\documentclass[fleqn,10pt,twocolumn]{wlscirep}
\usepackage[utf8]{inputenc}
\usepackage[T1]{fontenc}
\usepackage{bm}

\usepackage{booktabs}
\usepackage{adjustbox}
\usepackage{longtable}
\usepackage{caption}
\usepackage{mathtools}
\usepackage{amssymb}
\usepackage{amsmath}
\usepackage{etoolbox}
\usepackage{amsthm}
\usepackage{tocloft}

\allowdisplaybreaks

\title{Evolutionary dynamics in stochastic nonlinear public goods games}

\author[1,2,3,11]{Wenqiang Zhu}

\author[1,2,3,5,6,11]{Xin Wang}
\author[9]{Chaoqian Wang}
\author[1,2,3,5,6]{Longzhao Liu}
\author[10]{Jiaxin Hu}
\author[1,2,3,5,6,7,8]{Zhiming Zheng}
\author[1,2,3,4,5,6,7,8]{Shaoting Tang}
\author[4,*]{Hongwei Zheng}
\author[4,*]{Jin Dong}

\affil[1]{School of Artificial Intelligence, Beihang University, Beijing, 100191, China}
\affil[2]{Key laboratory of Mathematics, Informatics and Behavioral Semantics, Beihang University, Beijing 100191, China}
\affil[3]{Zhongguancun Laboratory, Beijing 100094, China}
\affil[4]{Beijing Academy of Blockchain and Edge Computing, Beijing 100085, China}
\affil[5]{Beijing Advanced Innovation Center for Future Blockchain and Privacy Computing, Beihang University, Beijing 100191, China}
\affil[6]{State Key Lab of Software Development Environment, Beihang University, Beijing 100191, China}
\affil[7]{Institute of Medical Artificial Intelligence, Binzhou Medical University, Yantai 264003, China}
\affil[8]{School of Mathematical Sciences, Dalian University of Technology, Dalian 116024, China}
\affil[9]{Department of Computational and Data Sciences, George Mason University, Fairfax, VA 22030, USA}
\affil[10]{Faculty of Business, The Hong Kong Polytechnic University, Hung Hom, Kowloon, Hong Kong 999077, China}
\affil[11]{These authors contributed equally: Wenqiang Zhu, Xin Wang}

\affil[*]{e-mail: dongjin@baec.org.cn~(J.~Dong), hwzheng@pku.edu.cn~(H.~zheng)}

\begin{abstract}
Understanding the evolution of cooperation in multiplayer games is of vital significance for natural and social systems. An important challenge is that group interactions often leads to nonlinear synergistic effects. However, previous models mainly focus on deterministic nonlinearity where the arise of synergy or discounting effect is determined by certain conditions, ignoring uncertainty and stochasticity in real-world systems. Here, we develop a probabilistic framework to study the cooperative behavior in stochastic nonlinear public goods games. Through both analytical treatment and Monte Carlo simulations, we provide comprehensive understanding of social dilemmas with stochastic nonlinearity in both well-mixed and structured populations. We find that increasing the degree of nonlinearity makes synergy more advantageous when competing with discounting, thereby promoting cooperation. Interestingly, we show that network reciprocity loses effectiveness when the probability of synergy is small. Moreover, group size exhibits nonlinear effects on group cooperation regardless of the underlying structure. Our findings thus provide novel insights into how stochastic nonlinearity influences the emergence of prosocial behavior.
\end{abstract}
\begin{document}

\flushbottom
\maketitle

\maketitle

\section*{}
The evolution of cooperative behavior among unrelated individuals presents a complex and significant challenge in both natural and human social systems~\cite{rand2013human,axelrod1981,pennisi2005}, as conflicts between individual and collective interest often lead to social dilemmas~\cite{van2013psychology}. Fortunately, evolutionary game theory~\cite{smith1973logic,hilbe2023evolutionary,weibull1997evolutionary,nowak2006evolutionary,lieberman2005evolutionary,allen2017evolutionary} provides a powerful tool for understanding the widespread cooperative behaviour in the real world~\cite{johnson2003puzzle,nelson2006}. Beginning from two-player games, multiple mechanisms that could facilitate the emergence and sustainability of cooperation have been identified in recent decades~\cite{nowak2006five,perc2010coevolutionary,zhu2022exposure,wang2024evolutionary}. For instance, group selection~\cite{traulsen2006evolution}, direct and indirect reciprocity~\cite{nowak1998evolution,van2012direct,schmid2021unified,romano2022direct,xia2023reputation}, punishment and reward~\cite{sigmund2001reward,perc2017statistical,zhu2023effects}, and importantly, spatial and network reciprocity~\cite{nowak1992evolutionary,nowak2010evolutionary,su2019spatial}.

When it comes to multiplayer interactions, the public goods game (PGG)~\cite{hauert2002volunteering,santos2008social,wang2022reversed} naturally extends the Prisoner's Dilemma to multi-player circumstance. In the classic PGG, group members decide whether to incur costs by contributing to a common pool. Each investment yields specific benefits, and the pool's total profits are evenly distributed among all players. Consequently, maximum benefit for each participant is achieved through full cooperation. However, the incentive to defect for higher individual gains persists, leading to the "tragedy of the commons"~\cite{hardin1968tragedy}. A natural question is, what are the essential differences between multi-player games and two-player games apart from the game structure? Recent advances have addressed this fundamental problem from different perspectives, such as the group-level reputation structure~\cite{wei2023indirect}, asymmetrical feedback from the environment~\cite{wang2020eco,wang2020steering}, heterogeneity of group members~\cite{mcavoy2020social}, and the underlying topological structures~\cite{battiston2020networks,alvarez2021evolutionary,sheng2024strategy,yagoobi2023categorizing}.

Note that so far, the above PGG assumes that each cooperator's input cost obtains an equivalent payoff, resulting in a linear group payoff structure. In reality, however, group interactions often exhibit nonlinear benefits, where the ability as well as the payoff of the group could be synergistically enhanced or reduced~\cite{archetti2012game,wu2016evolving,kristensen2022ancestral}. Hauert~{\it et~al.}~\cite{hauert2006synergy} first introduced the concepts of synergy and discounting into PGG to describe the  nonlinear payoff structure. This nonlinear mechanism mirrors economies of scale or diminishing returns, prevalent in group interactions within biological and social systems~\cite{boyd2007narrow,pena2015evolutionary}. For instance, in enzyme catalysis and regulation~\cite{hammes2012enzyme}, considering the enzyme as the public good, a given substrate concentration coupled with an increased enzyme concentration can result in efficiency that surpasses linear growth, indicative of synergy. Conversely, treating the substrate as the public good, a given enzyme concentration can facilitate a faster reaction with a smaller substrate amount, but the reaction rate approaches a limit with further substrate increases, demonstrating a discounting effect. Within this framework, many studies investigate the impacts of nonlinear interactions on the evolution of cooperation under multiple mechanisms, including spatial effects~\cite{hauert2006spatial,li2015evolutionary,li2014cooperation}, group-size diversity~\cite{pena2012group},  reputation~\cite{zhu2024reputation}, population dynamics~\cite{zhou2018coevolution,zhou2015evolution} and external environmental factors~\cite{jiang2023nonlinear,ma2023evolution,quan2024cooperation}.

Despite the progress, previous researches largely focused on deterministic nonlinearity where either synergy or discounting will appear deterministically under given conditions~\cite{hauert2006synergy,li2014cooperation}, overlooking the uncertainty and stochasticity in complex systems~\cite{liu2021co,wang2024evolution,hilbe2018evolution,su2019evolutionary}. For instance, in team sports such as basketball and football, a player cannot predict whether joining a team will enhance or weaken the team's performance before actually joining. Even star players cannot guarantee a synergistic effect and team success. For example, in the history of NBA,  the "Big Three" model can lead to success, such as when LeBron James, Dwyane Wade, and Chris Bosh joined forces on the Miami Heat, winning two championships. However, there are also many instances where this model has failed. The Los Angeles Lakers' attempt with Karl Malone, Gary Payton, Shaquille O'Neal, and Kobe Bryant in the 2003-2004 season did not result in a championship. Similarly, the Brooklyn Nets' recent assembly of Kevin Durant, James Harden, and Kyrie Irving did not lead to the expected success. Similar stochastic nonlinearity can also be observed in biological, economic and social systems~\cite{hammes2012enzyme,zhou2015evolution,kuchler2016enzymatic,quan2021effects}, where synergy and discounting coexist. In general, before joining the game, the participant does not know whether the complex group interactions will ultimately result in synergy or discounting effect. Notably, how such stochasticity in nonlinear PGG affects the evolutionary outcomes of cooperation remains largely unknown.

To fill this gap, we introduce a simple probability-based model to study the evolutionary dynamics of stochastic nonlinear PGG. We assume that each group has a certain probability of being either synergy or discounting. For mathematical simplicity, the model employs a uniform nonlinear coefficient to quantify the degree of nonlinearity. For a comprehensive understanding, we theoretically derive the evolutionary equations~\cite{taylor1978evolutionary,traulsen2006stochastic,ohtsuki2006simple} in both well-mixed and structured populations, the accuracy and robustness of which have been verified by Monte Carlo simulations. Our model recovers five possible evolutionary states: (\romannumeral 1) full cooperation, (\romannumeral 2) full defection, (\romannumeral 3) coexistence of cooperation and defection, (\romannumeral 4) bi-stable state of full cooperation and full defection, and (\romannumeral 5) bi-stable state of coexistence and full cooperation. We show that although an increase in the degree of nonlinearity enhances both synergy and discounting, synergy can still win the competition and thus promoting cooperation, even when the probability of synergy is smaller than that of discounting. Interestingly, we find that network reciprocity is not always satisfied, which is different from the deterministic situations. When the proportion of synergy group is relatively small, network structure exhibits inhibiting effect on cooperation compared to the well-mixed population. Further, we find the nonlinear influence of group size on the evolution of cooperation in all cases, where a moderate group size inhibits cooperation due to the combined influence of social dilemma and aggregation dilemma, while a large group size helps breaking the dilemmas by leveraging the increasing advantage of synergy. Our results provide important insights into the nonnegligible influence of stochasticity on cooperation behaviors in nonlinear systems.

\section*{Results}\label{sec_model}

\subsection*{Model overview}
In a PGG, player \(i\) has two strategies to choose from, namely cooperation (\(S_i=C\)) and defection (\(S_i=D\)). Every cooperator in the group contributes a cost \(c\) (\(c=1\)) to the common pool, while defectors contribute nothing. The total contribution in the common pool is multiplied by an enhancement factor \(r\) and then evenly distributed among all players in the group. The described PGG represents a linear payoff scenario, wherein each cooperator's contribution is identical. Here, we introduce a synergy and discounting factor \(\omega\), which assumes that each additional cooperator contributes progressively more (synergy) or less (discounting) to the collective benefits~\cite{hauert2006synergy}. When the group consists of \(G\) individuals, including $n_{C} > 0$ cooperators, the payoffs for cooperators and defectors are expressed as
\begin{subequations}\label{eq_1}
	\begin{align}
		\pi_D(n_{C})&=\frac {r}{G}(1+\omega+\omega^2+\cdots+\omega^{n_{C}-1})=\frac {r}{G}\frac{1-\omega^{n_{C}}}{1-\omega}, \\
		\pi_C(n_{C}) &=\pi_D(n_{C})-1, 
	\end{align}
\end{subequations}
and $\pi_D(0)=0$ if $n_{C}=0$. When \(w > 1\), the payoff calculation in Eq.~(\ref{eq_1}) reflects a synergy effect, where each additional cooperator brings more benefits. When \(w < 1\), it demonstrates a discounting effect, as each additional cooperator contributes less benefit. When \(w = 1\), Eq.~(\ref{eq_1}) degenerates into the linear PGG. 

\begin{figure*}[ht]
\centering
\includegraphics[width=0.88\textwidth]{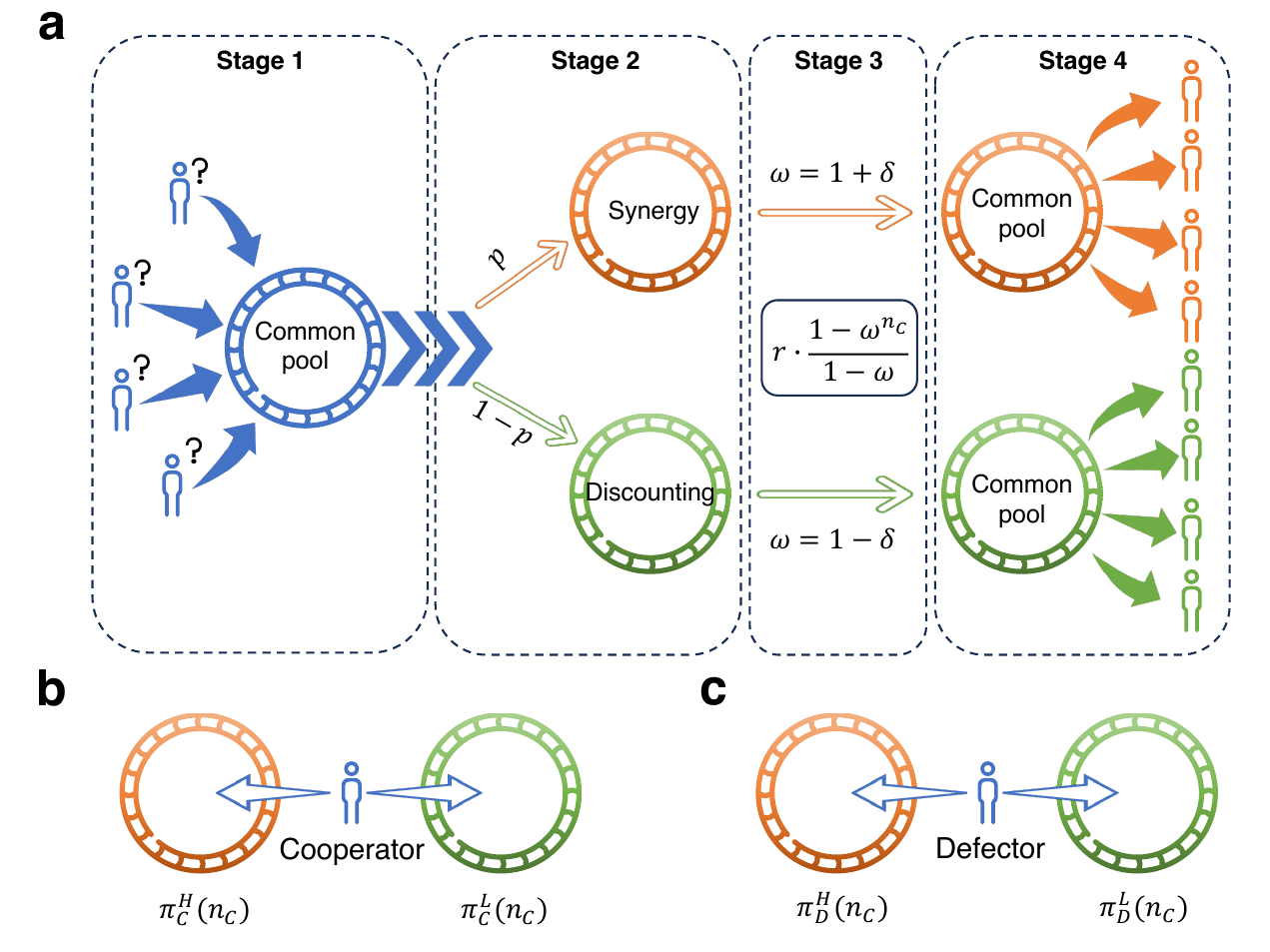}
\caption{\textbf{Schematic representation of the model.} \textbf{a} Process for calculating payoffs. Players decide whether to contribute to the common pool: cooperators contribute a cost, while defectors contribute nothing. The cumulative contribution in the common pool is nonlinear, with a probability $p$ to be a synergy effect and $1-p$ to be a discounting effect. When a synergy effect occurs, each additional contribution yields greater payoffs than the previous. Conversely, with the discounting effect, each additional contribution yields less payoffs. Finally, all benefits in the common pool are distributed equally among all players in the group. \textbf{b}, \textbf{c} Four payoff types: cooperation or defection in synergy or discounting groups.}\label{model}
\end{figure*}

\begin{figure*}[ht]
    \centering
    \includegraphics[width=1\textwidth]{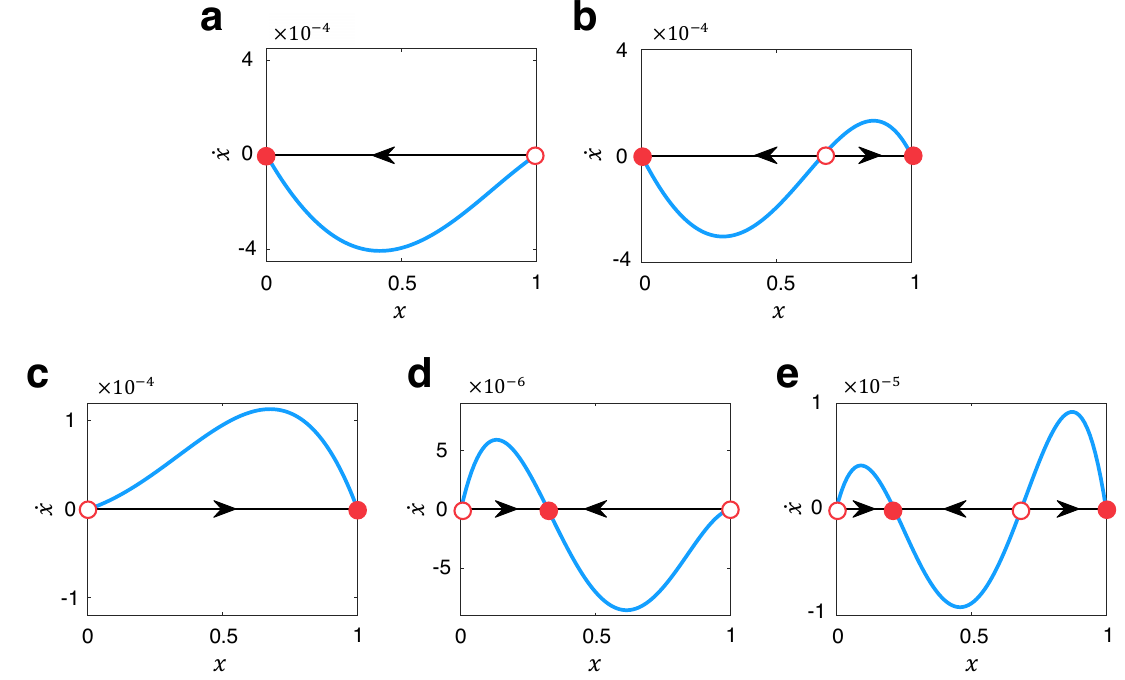}
    \caption{\textbf{Five evolutionary states in the population.} Solid circles denote stable equilibrium points, hollow circles denote unstable equilibrium points, and arrows denote evolutionary directions. Panels \textbf{a} and \textbf{b} represent the cases where $r<G$ for the two states of full defection, bi-stability of cooperation $\&$ defection, respectively. Panels \textbf{c}, \textbf{d}, and \textbf{e} represent the cases where $r>G$ for the three states of full cooperation, coexistence, bi-stability of coexistence $\&$ cooperation, respectively. The parameters are set as \textbf{a} $r=3$, $G=5$, $p=0.6$, $\delta=0.2$, \textbf{b} $r=3$, $G=5$, $p=0.6$, $\delta=0.4$, \textbf{c} $r=5.1$, $G=5$, $p=0.6$, $\delta=0.1$, \textbf{d} $r=5.1$, $G=5$, $p=0.4$, $\delta=0.1$, \textbf{e} $r=5.1$, $G=5$, $p=0.4$, $\delta=0.15$. }\label{dynamic}
\end{figure*}

While prior studies assumed that all common pools exhibit either a synergy or a discounting effect~\cite{hauert2006synergy}, here we explore scenarios in which synergy and discounting effects coexist in the population. Specifically, each group has a chance \(p\) of being a synergy group, and a chance \(1-p\) to be a discounting group. To distinguish between these two types, a nonlinear coefficient $\delta$ ($\delta \geq 0$) is introduced. In the synergy group, we set \(\omega=1 + \delta\), and in the discounting group, \(\omega=1 - \delta\). After such a nonlinear interaction, all benefits accumulated in the common pool are equally distributed to all players in the group. Fig.~\ref{model}\textbf{a} provides a schematic representation of the model.

Based on the above assumptions, various player types receive payoffs from gaming groups that exhibit varying effects. We label the group demonstrating synergy as high contribution ($H$) and the group illustrating discounting as low contribution ($L$). Let \(\pi_i^j\) represent the payoffs for strategy $i \in \{C,D\}$ participating in a public goods game with effect \(j \in \{H,L\}\). Figs.~\ref{model}\textbf{b} and \textbf{c} depict the scenarios for cooperators and defectors, respectively, in gaming groups with these two effects. Considering $\omega=1\pm \delta$, the payoffs for these scenarios are expressed as
\begin{subequations}\label{eq_2}
    \begin{align}
    \pi_{C}^{H}(n_{C}) &=\frac{r}{G}\cdot\frac{(1+\delta)^{n_{C}}-1}{\delta}-1,  \\
    \pi_{C}^{L}(n_{C}) &=\frac{r}{G}\cdot\frac{1-(1-\delta)^{n_{C}}}{\delta}-1,  \\
    \pi_{D}^{H}(n_{C}) &=\frac{r}{G}\cdot\frac{(1+\delta)^{n_{C}}-1}{\delta},  \\
    \pi_{D}^{L}(n_{C}) &=\frac{r}{G}\cdot\frac{1-(1-\delta)^{n_{C}}}{\delta}.
\end{align}
\end{subequations}

We assume that cooperator and defector have the same probability $p$ and $1-p$ of being located in the synergy and discounting groups. Moreover, the property of a group (synergy or discounting) is stochastic when a player calculates the payoffs from this group each time. In this way, the payoffs for cooperators and defectors are expressed as
\begin{subequations}\label{eq_3}
	\begin{align}
		\pi_{C}(n_{C})&=p\pi_{C}^{H}(n_{C})+(1-p)\pi_{C}^{L}(n_{C})\nonumber\\&=p\frac{r[(1+\delta)^{n_{C}}-1]}{G\delta}+(1-p)\frac{r[1-(1-\delta)^{n_{C}}]}{G\delta}-1,\\
		\pi_{D}(n_{C})&=p\pi_{D}^{H}(n_{C})+(1-p)\pi_{D}^{L}(n_{C})\nonumber\\&=p\frac{r[(1+\delta)^{n_{C}}-1]}{G\delta}+(1-p)\frac{r[1-(1-\delta)^{n_{C}}]}{G\delta},
	\end{align}
\end{subequations}
which still satisfies $\pi_C(n_{C})=\pi_D(n_{C})-1$.

In~\ref{sup3}, we provide a discussion of the general case (i.e., asymmetric nonlinear factors) and find that to a large extent the use of uniform nonlinear factors is representative of the evolutionary outcome of the system.

\subsection*{Evolutionary dynamics in well-mixed populations}
In well-mixed populations, players interact with each other with equal probability. Let $x$ represent the fraction of cooperators and $1-x$ the fraction of defectors. We can get the average payoff for cooperators ${\Pi}_C$ and defectors ${\Pi}_D$ as
\begin{subequations}\label{eq_Pi_well}
	\begin{align}
		\Pi_{C}&=\frac{r}{G\delta}[1-2p+p(1+\delta)(1+\delta         x)^{G-1}\nonumber\\
  &-(1-p)(1-\delta)(1-\delta x)^{G-1}]-1,\\
		\Pi_{D}&=\frac{r}{G\delta}[1-2p+p(1+\delta x)^{G-1}\nonumber\\
  &-(1-p)(1-\delta x)^{G-1}].
		\end{align}
\end{subequations}
Then the probability of a cooperator turning into defection is
\begin{equation}\label{eq_fermi}
    P_{C\leftarrow D}=\frac{1}{1+\mathrm{e}^{s(\Pi_{C}-\Pi_{D})}},
\end{equation}
where $s>0$ denotes the strength of selection. The higher this parameter is, the more likely one imitates the higher payoff players.

For arbitrary strength of selection and large populations~\cite{zhou2018coevolution}, the evolutionary dynamics can be expressed as
\begin{equation}\label{eq_dynamic_well}
	\begin{aligned}
		\dot{x}&=x(1-x)\tanh{\frac{sf(x)}{2}},
	\end{aligned}
\end{equation}
where $f(x)=\frac{r}{G}[p(1+\delta x)^{G-1}+(1-p)(1-\delta x)^{G-1}]-1$, which determines the sign of the hyperbolic tangent and hence the direction of evolution. When $\delta=0$, then $f(x)=r/G-1$, representing the linear PGG. If $r=G$, the population exhibits neutral drift; if $r>G$, cooperation emerges; if $r<G$, defection dominates. Nonlinear game dynamics emerges when $\delta >0$.

The system exhibits five distinct evolutionary states for different parameter values when $\delta >0$ (see~\ref{sup1} for theoretical analysis). Fig.~\ref{dynamic} numerically illustrates the five evolutionary states in well-mixed populations. For $r<G$, the population exhibits two distinct states. If $r[p\left(1+\delta\right)^{G-1}+\left(1-p\right)\left(1-\delta\right)^{G-1}]<G$, the population evolves to full defection, as shown in Fig.~\ref{dynamic}\textbf{a}. If $r[p\left(1+\delta\right)^{G-1}+\left(1-p\right)\left(1-\delta\right)^{G-1}]>G$, the population will evolve to bi-stability of cooperation and defection, characterized by an unstable internal equilibrium. The ultimate state depends on the initial proportion of cooperators; below this point, the population tends towards defection, and above it, towards cooperation, as depicted in Fig.~\ref{dynamic}\textbf{b}. For $r>G$, three distinct states are observed in the population. In this scenario, if the proportion of synergy groups equals or exceeds that of discounting groups (i.e., $p\geq0.5$), the population may evolve into a state of full cooperation, as shown in Fig.~\ref{dynamic}\textbf{c}. When the proportion of synergy groups is lower than the discounting groups (i.e., $p<0.5$), three states emerge. If $r[p(1+\delta)^{G-1}+(1-p)(1-\delta)^{G-1}]<G$, the population evolves into a coexistence state of cooperation and defection, with a stable internal equilibrium, as shown in Fig.~\ref{dynamic}\textbf{d}. If $r[p(1+\delta)^{G-1}+(1-p)(1-\delta)^{G-1}]>G$ and $\delta>\delta^{*}$ with $f(x^{**})<0$ (see \ref{sup1}~Eq.~(\ref{eq_x**})), the population achieves bi-stability of coexistence and cooperation, featuring both stable and unstable internal equilibrium points, as represented in Fig.~\ref{dynamic}\textbf{e}. For other parameter configurations at $p<0.5$, the population evolves towards full cooperation. We also verify the robustness of our theoretical results through monte carlo simulations in \ref{sup5}~Fig.~\textbf{S}\ref{simu_time}.

\begin{figure*}[!ht]
    \centering
    \includegraphics[width=1\textwidth]{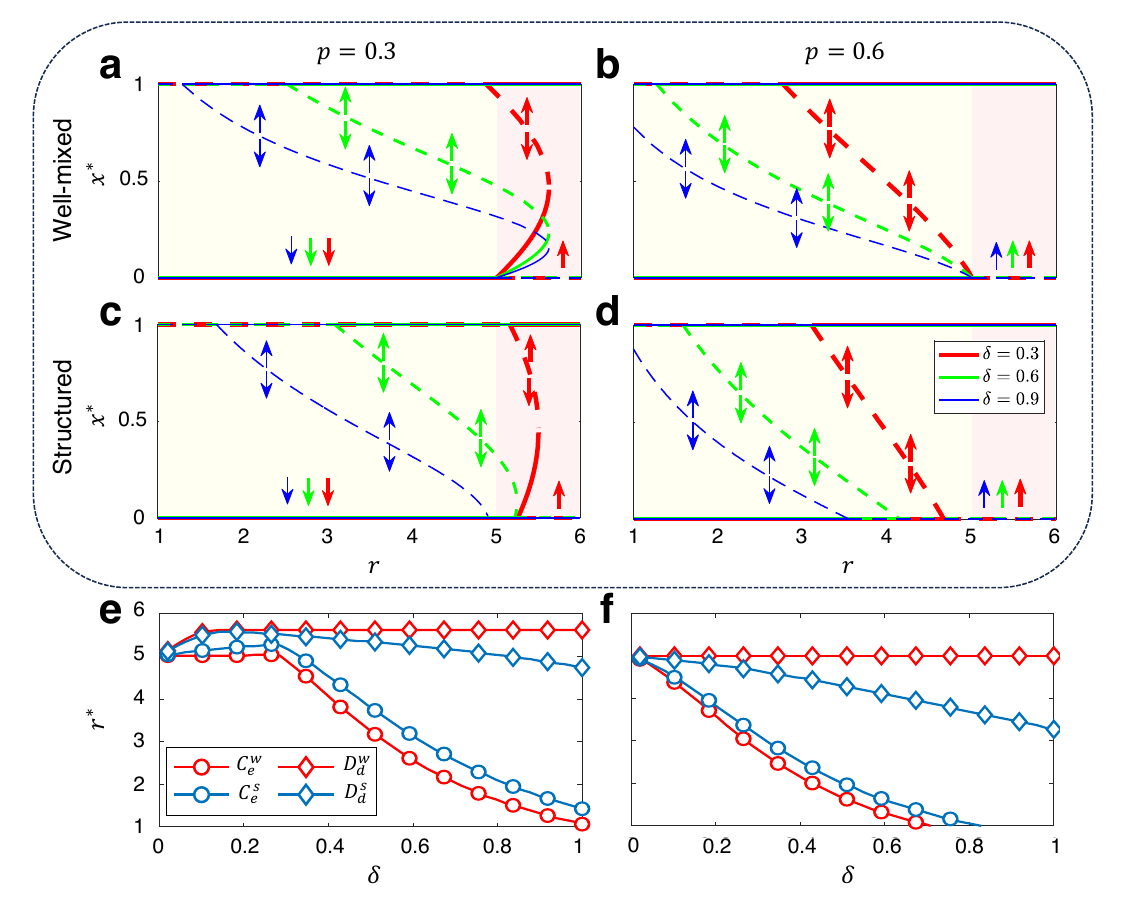}
    \caption{\textbf{Evolution of cooperation with stochastic nonlinear interactions.} Panels \textbf{a} to \textbf{d} show the equilibrium fraction of cooperators $x^{*}$ as a function of $r$ for different values of $\delta$. Every equilibrium is obtained by solving the equation $f(x)=0$ (well-mixed populations) and $g(p_C)=0$ (structured populations) numerically using the bisection method with the error 0.001. Solid lines indicate stable equilibrium, dashed lines indicate unstable equilibrium, and arrows indicate the direction of evolution. To facilitate comparison with linear PGG, set the background color of the region with $r<G$ (or $r < k + 1$) to yellowish and the other regions to pinkish. The increase in $\delta$ supports synergy as the dominant factor in evolution, which in turn promotes cooperation. Panels \textbf{e} and \textbf{f} show the critical $r$-values corresponding to the emergence of cooperators and the disappearance of defectors for low ($p=0.3$) and high ($p=0.6$) synergy group proportions, respectively. Where $C_e^{w}$ ($C_e^{s}$) represents the emergence of cooperators in well-mixed populations (structured populations) and $D_d^{w}$ ($D_d^{s}$) represents the disappearance of defectors in well-mixed populations (structured populations). It can be found that well-mixed populations are more conducive to the emergence of cooperation, while structured populations are more conducive to the stabilization of cooperation. The fixed parameters are $G=k+1=5$.}
	\label{hz_r}
\end{figure*}

\subsection*{Evolutionary dynamics in structured populations}
In evolutionary dynamics within structured populations, pair approximation~\cite{ohtsuki2006simple} proves to be an effective approach for describing the evolutionary process. In structured populations, players are situated on a regular graph with degree $k$. Each player engages in $k+1$ groups of public goods games centered on $k$ neighbors and itself.

At each time step, the same update rule is applied in order to be consistent with well-mixed populations. Let $p_C$ and $p_D$ denote the fraction of cooperators and defectors in the population. Under weak selection (i.e., $0<s\ll1$)~\cite{fu2009evolutionary,wang2023evolution,wang2023inertia}, we obtain the deterministic evolutionary dynamics in the structured population as
 \begin{equation}\label{eq_10}
	\dot{p}_C=\frac{s(k-2)}{4(k-1)}p_C(1-p_C)g(p_C),
\end{equation}
where $g(p_C)=\phi_1-\phi_2-\phi_3+\phi_4-(k-2)p_C(\phi_2-\phi_4)$ determines the direction of evolution. The derivation of $\dot{p}_C$ and the details of $\phi_1$ to $\phi_4$ are given in~\ref{sup2}.

\subsection*{The competition between synergy and discounting}
\begin{figure*}[!ht]
    \centering
    \includegraphics[width=1\textwidth]{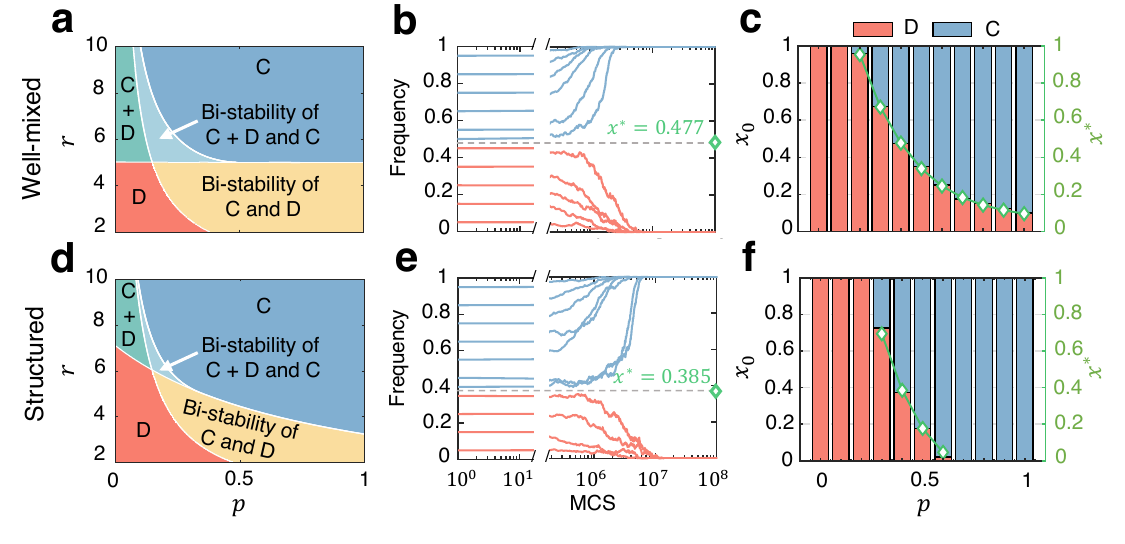}
    \caption{\textbf{Network structures do not always support cooperation.} Panels \textbf{a} and \textbf{d} show the phase diagrams of the r-p parameter plane when $\delta=0.6, G=k+1=5$. Panels \textbf{b} and \textbf{e} show the evolution of the frequency of the cooperator over time for different initial configurations when using Monte Carlo simulations when $p=0.4$. Panels \textbf{c} and \textbf{f} compare the simulation results with the theoretical results when $p \in [0,1]$. In \textbf{c} and \textbf{f}, the stacked plots corresponding to the left y-axis show the simulation results of the final population states corresponding to different initial cooperator frequencies, and the green curves corresponding to the right y-axis are the unstable internal equilibrium points in the theoretical results. It can be found that the simulation results are in good agreement with the theoretical results. The first row corresponds to the results for the well-mixed population and the second row corresponds to the results for the structured population. In Monte Carlo simulations, set parameters to $G=5$ and $s=0.1$ in well-mixed populations and $k=4$ and $s=0.02$ in structured populations. Other parameters were $N=1\times 10^4$, $r=4$, $\delta=0.6$}
	\label{simu_p}
\end{figure*}

The effects of synergy and discounting are related to the number of cooperators $n_C$ in the group and exhibit an asymmetric relationship compared to linear interactions. Specifically, as $n_C$ increases, the contribution from synergy tends toward $\infty$, while the contribution from discounting tends toward $1/ \delta$. Thus, increasing $\delta$ constrains the discounting effect, thereby expanding the advantage of synergy. The asymmetric relationship between the two nonlinear interactions prevents us from measuring the competition between the two simply through the parameter $p$. When these two conflicting effects coexist in a population, we consider the synergy to be dominant if the evolutionary outcome is better than the linear case, and vice versa for the discounting.

Based on the theoretical analysis about evolutionary dynamics, it is known that in the case of linear interactions, the system undergoes a phase transition from full defection to full cooperation at $r = G$ (or $r = k + 1$). Figs.~\ref{hz_r}\textbf{a, c} and \textbf{e} correspond to scenarios with a low proportion of synergy groups. When $\delta$ is small, a large $r$-value is needed for cooperation to emerge (especially for structured populations when $\delta\leq0.3$ ), indicating that discounting dominates the evolution of cooperation at this point. With further increases in $\delta$, synergy and discounting effects are increasing. However, due to the asymmetric relationship between the two, it is still possible for synergy to reverse and become a key factor dominating the evolution of cooperation (albeit with a low percentage of synergistic groups). It can be observed that the smallest $r$-value required to induce the emergence of cooperation is gradually decreasing. Moreover, the domain of attraction for full cooperation is gradually expanding, implying that cooperation is more favored. Therefore, an increase in $\delta$ facilitates synergy to dominate the evolution of cooperation. When the proportion of synergy group is large (see Figs.~\ref{hz_r}\textbf{b, d} and \textbf{e}), cooperation is promoted regardless of the value of $\delta$. This is indisputable because a high proportion of synergy groups unconditionally supports cooperation\cite{zhou2015evolution}. At this point synergy will play a far greater role in cooperation evolution than discounting. The phase diagram shown in Figs.~\ref{simu_p}\textbf{a} and \textbf{d} support the above conclusion, i.e., when $p>0.5$ cooperation is fully facilitated; when $p<0.5$, synergy still has the opportunity to dominate the evolution of cooperation thus facilitating the emergence of cooperation.

\subsection*{Network structures disfavor cooperation when synergy groups are few}

Network structure has been shown to be a key factor in facilitating the evolution of cooperation, however, this conclusion is not always met in our model. By comparing the emergence and stabilization of cooperation under the two populations, in Figs.~\ref{hz_r}\textbf{e} and \textbf{f} we find that the well-mixed population are more conducive to the emergence of cooperation (i.e., $C_e^{w}<C_e^{s}$), while structured populations are more conducive to the stabilization of cooperation (i.e., $D_d^{s}<D_d^{w}$). The former case is more pronounced when the proportion of synergy groups is low, and the latter case is pronounced when the proportion of synergy groups is high.

Turning our attention to the comparative analysis of these final evolutionary states, Figs.~\ref{simu_p}\textbf{a} and \textbf{d} provide the phase diagrams of the two type of populations in the full $r$-$p$ plane. In Fig.~\ref{simu_p}\textbf{a}, the dividing line $r = G = 5$ divides the parameter plane into two regions, with three states existing above the dividing line and two states existing below the dividing line. Conversely, in the structured population, this dividing line deviates from the horizontal state as $p$ increases and is presented with a negative slope. This leads to an expansion of the parameter region to which full defection belongs at lower $p$ values and an expansion of the parameter region to which full cooperation belongs at higher $p$ values. Thus, in contrast to well-mixed populations, when $p$ is small, population structure hinders the facilitating effect of synergy on cooperation and promotes the inhibiting effect of discounting on cooperation, and when $p$ is large, the opposite is true. From the full $r$-$p$ plane, structured populations have larger regions of full cooperation and full defection, inhibiting the coexistence and bi-stability of cooperation and defection.

Figs.~\ref{simu_p}\textbf{b, c, e} and \textbf{f} further validate our conclusions through Monte Carlo simulations. When the system exhibits a bi-stable cooperation \& defection state (see Fig.~\ref{simu_p}\textbf{b} and \textbf{e}), there exists a critical threshold. If the initial frequency of cooperators falls below this threshold, the population tends to evolve towards full defection, and conversely, towards full cooperation if it surpasses this critical value. The simulated critical value can be found to be consistent with the theoretical value of the unstable equilibrium point in the evolutionary dynamics. If that critical value is lower, it favors cooperation. By comparing the simulation results of the two types of populations, it can be found that the well-mixed population promotes cooperation better than the structured population at $p=0.2$ and $p=0.3$, but the structured population is more effective in promoting cooperation when $p>0.3$.

\begin{figure*}[!ht]
    \centering
    \includegraphics[width=1\textwidth]{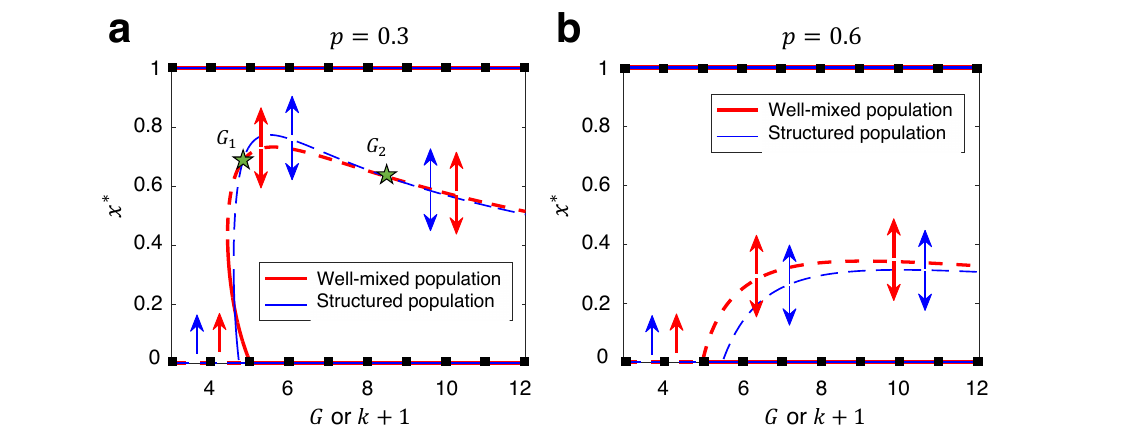}
    \caption{\textbf{Moderate group size inhibits the evolution of cooperation.} Panels \textbf{a} and \textbf{b} show the evolutionary dynamics for different group sizes $G$ (the group size of the well-mixed population) or $k+1$ (the group size of the structured population) when $p=0.3$ and $p=0.6$, respectively. In order to visualize the evolutionary results due to changes in group size, we continuousize the group size. The red line represents the well-mixed population, and the blue line represents the structured population. Dashed lines represent unstable equilibrium points, solid lines represent stable equilibrium points, and arrows represent the direction of evolution. The fixed parameters are $r=5$, $\delta=0.4$.}
	\label{hz_Gk}
\end{figure*}

When the proportion of synergy groups $p$ is low, the size of the group also influences the evolutionary results of the different populations. In Fig.~\ref{hz_Gk}\textbf{a}, the area between $G_1$ and $G_2$ reveals that moderate-sized groups in a well-mixed population have a larger basin of attraction for full cooperation compared to those in a structured population. Conversely, Fig.~\ref{hz_Gk}\textbf{b} shows that with a higher proportion of synergy groups, the structured population consistently surpasses the well-mixed population with the increase of group sizes.

In summary we can see that network structures do not always support cooperation and is like a double-edged sword. In our model, fewer synergy groups, compared to well-mixed populations, are the main factors driving the network structures to inhibit the evolution of cooperation. In addition, smaller enhancement factors and moderate group sizes exacerbated the inhibitory effect. When there are more synergy groups, however, the network structure was more conducive to the stabilization of cooperation.

\subsection*{Large group size promotes cooperation by utilizing the synergy effect}

When we focus on the impact of group size on the evolution of cooperation, as shown in Fig.~\ref{hz_Gk}, cooperation is initially promoted, then suppressed, and finally promoted again as group size increases. For a given PGG, cooperation is promoted in smaller groups because there is no social dilemma (or the intensity of the dilemma is very weak), allowing cooperation to prevail. As group size increases, cooperation is inhibited due to the dual factors of social dilemma and aggregation dilemma~\cite{roca2011emergence}. The aggregation dilemma exists because, on one hand, having more neighbors means participating in more PGGs, which can bring greater potential benefits. On the other hand, an increase in group size may reduce the payoff per contribution, leading to a reliance on others' contributions in large groups, thereby increasing risk.

In our model, the contribution in the group grows non-linearly with the number of cooperators, and the increase in the group size means that more cooperators can be accommodated. To measure the role of synergy and discounting in the aggregation dilemma, we introduce a metric AoS to represent the advantage of synergy:
\begin{equation}
    \mathrm{AoS}=p\vert \frac{(1+\delta)^{n_{C}}-1}{\delta} - n_C \vert - (1-p)\vert \frac{1-(1-\delta)^{n_{C}}}{\delta}-n_C \vert.
\end{equation}
If this metric is less than 0, it means that the additional income generated by synergy is less than the loss of income caused by discounting, which in turn leads to an intensification of the aggregation dilemma, and conversely, it helps to overcome the aggregation dilemma.

It can be found from Fig.~\textbf{S}\ref{competition_nc}\textbf{d} that as $n_C$ increases, the AoS first decreases and then increases. Therefore, for moderate $n_C$, the intensity of the aggregation dilemma is greater, and as $n_C$ further increases, the role of synergy gradually become significant, and at this time, synergy can transform the aggregation dilemma into aggregation advantages, thus promoting cooperation. Note that the above discussion only focuses on the aggregation dilemma. When we also consider the social dilemma, this will lead to the inhibition of cooperation under moderate group sizes, while large group sizes will promote the evolution of cooperation. This also provides insights into achieving large-scale group cooperation.

\section*{Discussion}
Nonlinear payoff structures are ubiquitous in natural and social systems. Previous research only examines either synergy or discounting in isolation~\cite{hauert2006synergy,hauert2006spatial,li2015evolutionary,li2014cooperation}, but these two effects usually coexist in reality~\cite{zhu2024reputation,zhou2018coevolution,zhou2015evolution}. On this basis, this paper considers the stochastic nonlinear interactions, where the payoff calculation of some groups follows the synergy effect and others reflect the discounting effect.

The coexistence of synergy and discounting in a population can affect the deterministic state of evolution. Notably, in contrast to the linear PGG, the presence of discounting groups implies that a well-mixed population may not invariably evolve to full cooperation, even with \(r>G\), particularly if the proportion of synergy groups is lower than that of discounting groups. Our theoretical results show that a total of five evolutionary states exist in the system, where the well-mixed population is strictly divided into two regions bounded by $r=G$. When $r<G$ two states exist, i.e., full defection and bi-stability of cooperation \& defection, and when $r>G$ three states exist, i.e., full cooperation, coexistence, and bi-stability of coexistence \& cooperation. Synergy promotes the evolution of cooperation, while discounting inhibits the evolution of cooperation, and these two conflicting interactions coexist and compete in the population. Compared with the linear PGG, if cooperation is promoted, we believe that synergy dominates the evolution of cooperation, otherwise it is a discount. Our results indicate that although an increase in the degree of nonlinearity enhances both synergy and discounting, synergy can still win the competition and thus promoting cooperation, even when the probability of synergy is smaller than that of discounting.

Since network reciprocity emerges as a critical mechanism in fostering cooperation, we also compare the evolutionary dynamics between well-mixed and structured populations. It emerges that structured populations do not universally outperform well-mixed populations~\cite{hauert2004spatial,li2016evolutionary}. Our results highlight that well-mixed populations more effectively facilitate the emergence of cooperation, whereas structured populations better support the stabilization of cooperation. From the full parameter plane, structured populations have larger regions of full cooperation and full defection, inhibiting the coexistence and bi-stability of cooperation and defection. Interestingly, we show that network structure is like a double-edged sword and disfavors cooperation when the probability of synergy is small. This implies that a structured framework can impede the effectiveness of a small number of synergy groups within the population. This conclusion is also validated by our simulation results.

Group size also plays an important role in evolutionary outcomes. Our results suggest that moderate number of neighbors are less conducive to the evolution of cooperation. As group size increases, a dual dilemma, the social dilemma and the aggregation dilemma~\cite{roca2011emergence}, emerges in the population. We measure aggregation dilemma by defining an advantage of synergy (AoS) metric and find that a moderate number of cooperators leads to the greatest aggregation dilemma strength. When we also take the social dilemma into account, this leads to suppressed cooperation behavior at moderate group sizes. As the group size increases further, AoS increases explosively, which will help overcome the aggregation dilemma and thus promote cooperation. Thus, synergy interaction enables large-scale group cooperation.

This paper presents a generalized scenario, offering a theoretical framework to study the stochastic nonlinear multiplayer game. However, the real world tends to be more complex, and it remains unexplored to consider probabilities based on environmental feedback~\cite{wang2020steering,wang2020eco} and group reputation~\cite{wei2023indirect} in future work. Moreover, our work only considers regular networks for structured populations, and it is of interest to study these two nonlinear interactions on more complex and real networks~\cite{li2020evolution,su2023strategy,wang2023public}. Synergy and discounting exemplify nonlinear interactions, which are essentially pairwise interactions, and it would be interesting to extend such nonlinear interactions to higher-order interactions~\cite{battiston2020networks,alvarez2021evolutionary,sheng2024strategy} as well. 

\section*{Methods}\label{Methods_sec}
Here, we provide methods for the evolutionary equations of well-mixed populations and structured populations when synergy interactions and discounting interactions coexist in a PGG.

\subsection*{Evolutionary equations of well-mixed populations}

In a well-mixed population comprising $N$ individuals with $i$ cooperators, the average payoffs for cooperators $\Pi_{C}$ and defectors $\Pi_{D}$ within a randomly sampled group of size $G$ can be expressed as
\begin{subequations}
	\begin{align}
		\Pi_C&=\sum_{k_C=0}^{G-1}\frac{\displaystyle{\binom{i-1}{k_C}\binom{N-i}{G-k_C-1}}}{\displaystyle{\binom{N-1}{G-1}}}\pi_C(k_C+1),\\
		\Pi_D&=\sum_{k_C=0}^{G-1}\frac{\displaystyle{\binom{i}{k_C}\binom{N-i-1}{G-k_C-1}}}{\displaystyle{\binom{N-1}{G-1}}}\pi_D(k_C).
	\end{align}
\end{subequations}

As $N \to \infty$, the binomial distribution can be used to approximate the hypergeometric distribution. We let $x=i/N$ denote the fraction of cooperators, then we get Eq.~(\ref{eq_Pi_well}). The strategy evolution following the asynchronous Fermi update rule~\cite{szabo1998evolutionary}, the probability of a cooperator turning into defection is Eq.~(\ref{eq_fermi}).

Excluding self-interaction and in the absence of mutations, the total number of individuals with a specific strategy changes by 1 only when the two selected individuals have differing strategies. The above process can be represented as a finite state Markov process with an associated tridiagonal transition matrix~\cite{traulsen2006stochastic}. Then the probability of increasing the number of cooperators from $i$ to $i + 1$ and the probability of decreasing from $i$ to $i-1$ are described as
\begin{equation}
    T_{i}^{\pm}=\frac{i}{N}\frac{N-i}{N}\frac{1}{1+\mathrm{e}^{{\mp s(\Pi_{C}-\Pi_{D})}}}.
\end{equation}
Thus, for arbitrary strength of selection and large populations, the evolutionary dynamics equation can be expressed as Eq.~(\ref{eq_dynamic_well}).

We focus on the evolutionary results when the population reaches stability. By setting $\dot{x} = 0$, we determine the equilibrium points. The stability of these equilibria is assessed by evaluating the evolutionary direction of the points: a negative slope indicates stability, whereas a positive slope suggests instability. \ref{sup1} theoretically gives the evolutionary state of the system and its stability conditions when synergies and discounts coexist.

\subsection*{Evolutionary equations of structured populations}

In a structured population comprising $N$ individuals, players are positioned on a regular graph with degree $k$, where the size of each group is $k + 1$. For a randomly selected individual $i$ (with a payoff of $\Pi_i$) and a randomly selected neighbor $j$ (with a payoff of $\Pi_j$), the probability that player $i$ imitates the strategy of neighbor $j$ under weak selection (i.e., $0<s\ll1$)~\cite{fu2009evolutionary,wang2023evolution,wang2023inertia} can be expanded as
\begin{equation}\label{eq_9}
	P_{i\leftarrow j}=\frac{1}{2}+\frac{\Pi_j-\Pi_i}{4}s+\mathrm{O}\left(s^2\right).
\end{equation}

Let $p_C$ and $p_D$ denote the fraction of cooperators and defectors in the population, then $p_C+p_D=1$. Let $q_{D|C}$ denote the conditional probability to find a defector given that the adjacent node is occupied by a cooperator, then $q_{C|C}+q_{D|C}=1$. Let $p_{CD}$ denote the frequencies of CD pairs, then we get $p_{CD}=p_{DC}=p_Dq_{C|D}=p_Cq_{D|C}$. At this point, the whole system can be described by only $p_C$ and $q_{C|C}$, where
\begin{subequations}
    \begin{align}
        p_D&=1-p_C,\\
        q_{D|C}&=1-q_{C|C},\\ p_{CD}&=p_{DC}=p_C\left(1-q_{C|C}\right),\\ q_{C|D}&=\frac{p_C\left(1-q_{C|C}\right)}{1-p_C},\\ q_{D|D}&=\frac{1-2p_C+p_Cq_{C|C}}{1-p_C},\\
        p_{DD}&=1-2p_C+p_Cq_{C|C}.
		\end{align}
\end{subequations}

We consider a focal cooperator replaced by a neighborhood of defectors ${P}_{C\leftarrow D}$ and a focal defector replaced by a neighborhood of cooperators ${P}_{D\leftarrow C}$, respectively. Assuming each player can undertake strategic replacement within a unit of time, the closed dynamics system is expressed as
 \begin{subequations}
	\begin{align}
		{\dot{p}}_C=&~\frac{p_{D}}{N}\sum_{k_C=0}^{k}{\binom{k}{k_C}q_{C|D}^{k_C}q_{D|D}^{k-k_C}{P}_{D\leftarrow C}}\nonumber\\
            &-\frac{p_{C}}{N}\sum_{k_C=0}^{k}{\binom{k}{k_C}q_{C|C}^{k_C}q_{D|C}^{k-k_C}{P}_{C\leftarrow D}},\\
		{\dot{p}}_{CC}=&\sum_{k_C=0}^{k}{\frac{2k_C}{kN}p_D\binom{k}{k_C}q_{C|D}^{k_C}q_{D|D}^{k-k_C}{P}_{D\leftarrow C}}\nonumber\\
            &-\sum_{k_C=0}^{k}{\frac{2k_C}{kN}p_C\binom{k}{k_C}q_{C|C}^{k_C}q_{D|C}^{k-k_C}{P}_{C\leftarrow D}}.
	\end{align}
\end{subequations}

The specific forms of ${\dot{p}}_C$ and ${\dot{p}}_{CC}$ are given in \ref{sup2}, it can be found that $q_{C|C}$ equilibrates much more faster than $p_C$ when $s$ is small and $s{\dot{p}}_{CC} \neq 0$. Thus, the system can be described only by $p_C$. When $N$ is large ($N \to \infty$), the dynamical equations can be expressed as Eq.~(\ref{eq_10}).

Similar to the well-mixed population, we can solve the equation $\dot{p}_C=0$ and find the equilibrium points numerically.

\section*{Data availability}
All data generated or analysed during this study are included within the paper and its supplementary information files.

\section*{Code availability}
The codes are written using MATLAB R2020a and Python 3.9.16. All source codes related to the work can be found at \url{https://github.com/JoeWynn7/Stochastic-PGG}

\bibliography{ms}

\section*{Acknowledgements}
This work is supported by National Science and Technology Major Project (2022ZD0116800), Program of National Natural Science Foundation of China (62141605, 12201026, 12301305), and Beijing Natural Science Foundation (Z230001).

\section*{Author contributions}
W.Z. and X.W. conceived of the study. W.Z., X.W. and C.W. designed the methodology. W.Z., X.W., C.W., L.L., J.H., Z.Z., S.T., H.Z., J.D. contributed and implemented the investigation. W.Z., X.W. and C.W. implemented the visualization. W.Z., X.W., C.W. and J.D. wrote and edited the manuscript.

\section*{Competing interests}
The authors declare no competing interests.

\section*{Additional information}
\textbf{Supplementary information} Supplementary information accompanies this paper at url will be inserted by publisher.

\section*{Correspondence}
Correspondence and requests for materials should be addressed to J. Dong.~(email: \href{mailto:dongjin@baec.org.cn}{dongjin@baec.org.cn}) and H. Zheng.~(email: \href{hwzheng@pku.edu.cn}{hwzheng@pku.edu.cn}).

\newpage
\pagebreak
\clearpage
\onecolumn
\setcounter{equation}{0}
\setcounter{figure}{0}
\setcounter{table}{0}
\setcounter{section}{0}
\makeatletter
\renewcommand{\thefigure}{\arabic{figure}}
\renewcommand{\figurename}{{\bf Supplementary Figure.}}
\renewcommand{\theequation}{S\arabic{equation}}
\renewcommand{\thesection}{{{\bf Supplementary Note \arabic{section}}}}
\renewcommand{\thesubsection}{\arabic{subsection}}

\newcounter{fnnumber}
\renewcommand{\thefootnote}{\fnsymbol{footnote}}
\begin{center}
{\Huge ---Supplementary Material---}\\[1em]
\textbf{\huge
Evolutionary dynamics in stochastic nonlinear public goods games}\\
\vspace{0.5cm}
Wenqiang Zhu$^{1,2,3,11}$, Xin Wang$^{1,2,3,5,6,11}$, Chaoqian Wang$^{9}$, Longzhao Liu$^{1,2,3,5,6}$, Jiaxin Hu$^{10}$, \\Zhiming Zheng$^{1,2,3,5,6,7,8}$, Shaoting Tang$^{1,2,3,5,6,7,8}$, Hongwei Zheng$^{4,*}$, Jin Dong$^{4,*}$\\
\vspace{0.5cm}

$^{1}${\it School of Artificial Intelligence, Beihang University, Beijing, 100191, China}\\
$^{2}${\it Key laboratory of Mathematics, Informatics and Behavioral Semantics, Beihang University, Beijing 100191, China}\\
$^{3}${\it Zhongguancun Laboratory, Beijing 100094, China}\\
$^{4}${\it Beijing Academy of Blockchain and Edge Computing, Beijing 100085, China}\\
$^{5}${\it Beijing Advanced Innovation Center for Future Blockchain and Privacy Computing, Beihang University, Beijing 100191, China}\\
$^{6}${\it State Key Lab of Software Development Environment, Beihang University, Beijing 100191, China}\\
$^{7}${\it Institute of Medical Artificial Intelligence, Binzhou Medical University, Yantai 264003, China}\\
$^{8}${\it School of Mathematical Sciences, Dalian University of Technology, Dalian 116024, China}\\
$^{9}${\it Department of Computational and Data Sciences, George Mason University, Fairfax, VA 22030, USA}\\
$^{10}${\it Faculty of Business, The Hong Kong Polytechnic University, Hung Hom, Kowloon, Hong Kong 999077, China}\\
$^{11}${\it These authors contributed equally: Wenqiang Zhu, Xin Wang}\\

$^{*}${\it e-mail: dongjin@baec.org.cn~(J.~Dong), hwzheng@pku.edu.cn~(H.~zheng)}\\

\end{center}
\textbf{This PDF file includes:} 
\begin{itemize}
	\item \textbf{Supplementary Note 1.} Theoretical analysis in well-mixed populations. 
	\item \textbf{Supplementary Note 2.} Theoretical analysis in structured populations. 
	\item \textbf{Supplementary Note 3.} Asymmetric nonlinear coefficients in public goods games. 
 	\item \textbf{Supplementary Note 4.} Synergy and discounting in aggregation dilemma. 
        \item \textbf{Supplementary Note 5.} Monte Carlo simulation to validate evolutionary dynamics.
	\item \textbf{Supplementary Figure 1.} Synergy as a major determinant of the evolution of cooperation.
        \item \textbf{Supplementary Figure 2.} Synergy can solve the aggregation dilemma.
        \item \textbf{Supplementary Figure 3.}  Simulation results in well-mixed populations.

\end{itemize}

\newpage
\section{Theoretical analysis in well-mixed populations.}\label{sup1}
In this section, we analyze the evolutionary dynamics in well-mixed populations, giving possible states of the population and their stability conditions. 

In well-mixed populations, players interact with each other with equal probability. When $N \to \infty$, the selection of panelists can be approximated using the binomial distribution~\cite{SM_hauert2006}. Let $x$ denote the fraction of cooperators and $G$ denote the group size, the average payoffs of cooperators and defectors are
\begin{subequations}\label{eqs1}
\begin{align}
		\Pi_C&=\sum_{i=0}^{G-1}\binom{G-1}{i}x^i(1-x)^{G-1-i}\pi_C(i+1),\\
		\Pi_D&=\sum_{i=0}^{G-1}\binom{G-1}{i}x^i(1-x)^{G-1-i}\pi_D(i).
	\end{align}
\end{subequations}
Then we get
\begin{subequations}
	\begin{align}
		\Pi_{C}&=\frac{r}{G\delta}[1-2p+p(1+\delta)(1+\delta         x)^{G-1}-(1-p)(1-\delta)(1-\delta x)^{G-1}]-1,\\
		\Pi_{D}&=\frac{r}{G\delta}[1-2p+p(1+\delta x)^{G-1}-(1-p)(1-\delta x)^{G-1}].
		\end{align}
\end{subequations}
Define $f(x)=\Pi_{C}-\Pi_{D}=\frac{r}{G}[p(1+\delta x)^{G-1}+(1-p)(1-\delta x)^{G-1}]-1$, then $f^{\prime}(x)=\frac{r(G-1)\delta}{G}[p(1+\delta x)^{G-2}-(1-p)(1-\delta x)^{G-2}]$. Setting $f^\prime(x)=0$ yields 
\begin{equation}\label{eq_x**}
    x^{**}=\frac{1-{(\frac{p} {1-p})}^\frac{1}{G-2}}{\delta[1+{(\frac{p}{1-p})}^\frac{1}{G-2}]}.
\end{equation}
Thus, if $x<x^{**}$, then $f^\prime(x)<0$ and $f(x)$ decreases monotonically; if $x>x^{**}$, $f^\prime(x)>0$ and $f(x)$ increases monotonically. If ${(\frac{p}{1-p})}^\frac{1}{G-2}\geq1$ (i.e., $p\geq0.5$), then $x^{**}\le0$, and $f(x)$ increases monotonically over $(0,1)$. If ${(\frac{p}{1-p})}^\frac{1}{G-2}<1$ (i.e., $p<0.5$), then $x^{**}>0$. Here, $f(x)$ decreases and then increases within $(0,1)$ if $x^{**}<1$ (i.e., $\delta>\frac{1-{(\frac{p}{1-p})}^\frac{1}{G-2}}{1+{(\frac{p}{1-p})}^\frac{1}{G-2}}=\delta^*$), and decreases monotonically if $x^{**}>1$ (i.e., $\delta<\delta^*$). Additionally, $\lim\limits_{x\to0^+}{f(x)}=\frac{r}{G}-1$ and $\lim\limits_{x\to1^{-}}{f(x)}=\frac{r}{G}[p(1+\delta)^{G-1}+(1-p)(1-\delta)^{G-1}]-1$. In summary, it can be concluded that the system states are categorized as follows:
\begin{itemize}
	\item[1.] When $\lim\limits_{x\to0^+}{f(x)}<0$ (i.e., $r/G<1$), there is no need to determine whether $x^{**}$ is positive or negative and exist two states:
	\begin{itemize}
		\item[(1)] When $\lim\limits_{x\to1^-}{f(x)}<0$, that is, when $r[p\left(1+\delta\right)^{G-1}+\left(1-p\right)\left(1-\delta\right)^{G-1}]<G$, the population is in the stabilized state of full defection;
		\item[(2)] When $\lim\limits_{x\to1^-}{f(x)}>0$, that is, $r[p\left(1+\delta\right)^{G-1}+\left(1-p\right)\left(1-\delta\right)^{G-1}]>G$, at this point, the population reaches the bi-stability of cooperation and defection with an unstable internal equilibrium.
	\end{itemize}
\end{itemize}
\begin{itemize}
	\item[2.] When $\lim\limits_{x\to0^+}{f(x)}>0$ (i.e., $r/G>1$), it is necessary to determine the positivity or negativity of $x^{**}$ and there exist three states:
	\begin{itemize}
		\item[(1)]When $p\geq0.5$, $x^{**}<0$, there is no need to judge the state of $\lim\limits_{x\to1^-}{f(x)}$ and the population is in a full cooperation steady state;
		\item[(2)]When $p<0.5$, $x^{**}>0$, it is necessary to determine the state of $\lim\limits_{x\to1^-}{f(x)}$. Thus two aspects need to be considered. When $\lim\limits_{x\to1^-}{f(x)}<0$, meaning that $r[p\left(1+\delta\right)^{G-1}+\left(1-p\right)\left(1-\delta\right)^{G-1}]<G$, at which point cooperation and defection coexist in the population. There exists an internally stable equilibrium. When $\lim\limits_{x\to1^-}{f(x)}>0$, that is, $r[p\left(1+\delta\right)^{G-1}+\left(1-p\right)\left(1-\delta\right)^{G-1}]>G$. If $\delta<\delta^*$ or $\delta>\delta^*$ and $f(x^{**})>0$, there is no internal equilibrium and the population is in a full cooperation steady state; if $ \delta>\delta^*$ and $f(x^{**})<0$, the population reaches the bi-stability of coexistence and cooperation with a stable internal equilibrium and an unstable internal equilibrium. 
		
	\end{itemize}
\end{itemize}

In summary, there are five evolutionary states under different parameter regions, namely full defection, bi-stability of cooperation and defection, coexistence, bi-stability of coexistence and cooperation, and full cooperation.

\section{Theoretical analysis in structured populations.}\label{sup2}

In structured populations, pair approximation~\cite{SM_simplerule,SM_PW2,SM_PW3,SM_PW4} is adopted to capture the evolution of strategies. We consider a regular graph of size $N$ and degree $k$, and each focal player is connected to $k_C$ cooperators as well as $k_D$ defectors (with $k_C+k_D=k$). Thus the size of each population is $k + 1$ and each player participates in $k + 1$ rounds of the public goods game.

Initially, consider a focal cooperator is replaced by a defector neighborhood. The cumulative payoffs of the focal cooperator and the defector are represented as ${\Pi}_F^C$ and $\Pi_D^C$, respectively. Under weak selection~\cite{SM_zhou2017}, the probability that the focal cooperator will be replaced by a defector in the neighborhood is
 \begin{equation}\label{eq_b1}
 	{P}_{C\leftarrow D}=\frac{k_D}{k}\left[\frac{1}{2}+\frac{{\Pi}_D^C-{\Pi}_F^C}{4}s+\mathrm{O}\left(s^2\right)\right].
 \end{equation}

The accumulated payoffs of the focal cooperator and a defector neighborhood are
\begin{subequations}\label{eq_b2}
	\begin{align}
		{\Pi}_F^C&=\pi_C\left(k_C+1\right)+k_C\sum_{i=0}^{k-1}{\binom{k-1}{i}q_{C|C}^iq_{D|C}^{k-1-i}\pi_C\left(i+2\right)}
		+k_D\sum_{i=0}^{k-1}{\binom{k-1}{i}q_{C|D}^iq_{D|D}^{k-1-i}\pi_C\left(i+1\right)},\\
		{\Pi}_D^C&=\pi_D\left(k_C+1\right)+\sum_{i=0}^{k-1}{\binom{k-1}{i}q_{C|D}^iq_{D|D}^{k-1-i}\pi_D\left(i+1\right)}
		+\left(k-1\right)q_{C|D}\sum_{i=0}^{k-1}{\binom{k-1}{i}q_{C|C}^iq_{D|C}^{k-1-i}\pi_D\left(i+1\right)}\nonumber\\&+\left(k-1\right)q_{D|D}\sum_{i=0}^{k-1}{\binom{k-1}{i}q_{C|D}^iq_{D|D}^{k-1-i}\pi_D\left(i\right)}.
	\end{align}
\end{subequations}

It can be further obtained that
\begin{equation}\label{eq_b3}
	\begin{aligned}
		\Pi_D^C-\Pi_F^C &=k+1+\frac{r}{(k+1)\delta}\{(k-1)q_{C|D}(\lambda_1-\lambda_2-\lambda_3+\lambda_4)-\delta(k-1)(\lambda_2+\lambda_4)\\
        &-k_C[(1+\delta)(\lambda_1-\lambda_2)-\left(1-\delta\right)(\lambda_3-\lambda_4)]\},
	\end{aligned}
\end{equation}
where 
\begin{subequations}
	\begin{align}
		\lambda_1&=p(1+\delta)(1+q_{C|C}\delta)^{k-1},\\
            \lambda_{2}&=p(1+q_{C|D}\delta)^{k-1},\\
            \lambda_{3}&=(1-p)(1-\delta)\big(1-q_{C|C}\delta\big)^{k-1},\\
            \lambda_4&=(1-p)(1-q_{C|D}\delta)^{k-1}.
	\end{align}
\end{subequations}

Similarly, consider a scenario where a focal defector is supplanted by a cooperator neighborhood. The cumulative payoffs of the focal defector and the cooperator are represented as ${\Pi}_F^D$ and $\Pi_C^D$, respectively. Under weak selection, the probability that the focal defector will be replaced by a cooperator neighborhood is
  \begin{equation}\label{eq_b4}
 	{P}_{D\leftarrow C}=\frac{k_C}{k}\left[\frac{1}{2}+\frac{{\Pi}_C^D-{\Pi}_F^D}{4}s+\mathrm{O}\left(s^2\right)\right].
 \end{equation}

The accumulative payoffs of the focal defector and a cooperator neighborhood are 
 \begin{subequations}\label{eq_b5}
	\begin{align}
		{\Pi}_F^D&=\pi_D\left(k_C\right)+k_C\sum_{i=0}^{k-1}{\binom{k-1}{i}q_{C|C}^iq_{D|C}^{k-1-i}\pi_D\left(i+1\right)}
		+k_D\sum_{i=0}^{k-1}{\binom{k-1}{i}q_{C|D}^iq_{D|D}^{k-1-i}\pi_D\left(i\right)},\\
		{\Pi}_C^D&=\pi_C\left(k_C\right)+\sum_{i=0}^{k-1}{\binom{k-1}{i}q_{C|C}^iq_{D|C}^{k-1-i}\pi_C\left(i+1\right)}
		+\left(k-1\right)q_{C|C}\sum_{i=0}^{k-1}{\binom{k-1}{i}q_{C|C}^iq_{D|C}^{k-1-i}\pi_C\left(i+2\right)}\nonumber\\
		&+\left(k-1\right)q_{D|C}\sum_{i=0}^{k-1}{\binom{k-1}{i}q_{C|D}^iq_{D|D}^{k-1-i}\pi_C\left(i+1\right)}.
	\end{align}
\end{subequations}

Further we can get
\begin{equation}\label{eq_b6}
	\begin{aligned}
		{\Pi}_C^D-{\Pi}_F^D&=-\left(k+1\right)+\frac{r}{\left(k+1\right)\delta}\{\left(k-1\right)q_{C|C}[(1+\delta)(\lambda_1-\lambda_2)-\left(1-\delta\right)(\lambda_3-\lambda_4)]\\
        &+(\lambda_1-\lambda_2-\lambda_3+\lambda_4)+\delta(k-1)(\lambda_2+\lambda_4)-k_C(\lambda_1-\lambda_2-\lambda_3+\lambda_4)\}.
	\end{aligned}
\end{equation}

Assuming each player can undertake strategic replacement within a unit of time, the closed dynamics system is expressed as
\begin{subequations}\label{eq_b7}
	\begin{align}
		{\dot{p}}_C&=\frac{p_{D}}{N}\sum_{k_C=0}^{k}{\binom{k}{k_C}q_{C|D}^{k_C}q_{D|D}^{k-k_C}{P}_{D\leftarrow C}}
            -\frac{p_{C}}{N}\sum_{k_C=0}^{k}{\binom{k}{k_C}q_{C|C}^{k_C}q_{D|C}^{k-k_C}{P}_{C\leftarrow D}},\\
		{\dot{p}}_{CC}&=\sum_{k_C=0}^{k}{\frac{2k_C}{kN}p_D\binom{k}{k_C}q_{C|D}^{k_C}q_{D|D}^{k-k_C}{P}_{D\leftarrow C}}
            -\sum_{k_C=0}^{k}{\frac{2k_C}{kN}p_C\binom{k}{k_C}q_{C|C}^{k_C}q_{D|C}^{k-k_C}{P}_{C\leftarrow D}}.
	\end{align}
\end{subequations}

Furthermore, we have
\begin{subequations}\label{eq_b8}
	\begin{align}
		\dot{p}_{C}&=\frac{p_{D}}{4kN}s\sum_{k_{C}=0}^{k}\binom{k}       {k_{C}}q_{C|D}^{k_{C}}q_{D|D}^{k_{D}}k_{C}(\Pi_{C}^{D}-      \Pi_{F}^{D})
        -\frac{p_{C}}{4kN}s\sum_{k_{C}=0}^{k}\binom{k}{k_{C}}q_{C|C}^{k_{D}}q_{D|C}^{k_{D}}k_{D}(\Pi_{D}^{C}-\Pi_{F}^{C})+O(s^{2}), \\
		\dot{p}_{CC}&=\sum_{k_{C}=0}^{k}\frac{2k_{C}}{k}p_{D}\binom{k}{k_{C}}q_{C|D}^{k_{C}}q_{D|D}^{k_{D}}\frac{k_{C}}{2kN}-\sum_{k_{C}=0}^{k}\frac{2k_{C}}{kN}p_{C}\binom{k}{k_{C}}q_{C|C}^{k_{D}}q_{D|C}^{k_{D}}\frac{k_{D}}{2k}+O(s)\nonumber\\
		&=\frac{p_{CD}}{kN}[(k-1)q_{C|D}+1]-\frac{k-1}kp_{CD}q_{C|C}+O(s) \nonumber\\
		&=\frac{p_{CD}}{kN}[1+(k-1)(q_{C|D}-q_{C|C})]+O(s).
	\end{align}
\end{subequations}
Since $q_{C|C}=\frac{p_{CC}}{p_C}$, it follows that
\begin{equation}\label{eq_b9}
	{\dot{q}}_{C|C}=\frac{{\dot{p}}_{CC}}{p_C}=\frac{p_{CD}}{kNp_C}[1+(k-1)(q_{C|D}-q_{C|C})]+O(s).
\end{equation}
Thus ${\dot{p}}_C$ and ${\dot{q}}_{C|C}$ can be transformed as
\begin{subequations}\label{eq_b10}
	\begin{align}
		{\dot{p}}_C&=sF_1(p_C,\ q_{C|C})+O(s^2),\\
		{\dot{q}}_{C|C}&=F_2(p_C,\ q_{C|C})+O(s),
	\end{align}
\end{subequations}
where
\begin{subequations}\label{eq_b11}
	\begin{align}
		F_1(p_C,q_{C|C})&=\frac{1-p_{C}}{4kN}\sum_{k_{C}=0}^{k}\binom{k}{k_{C}}\left(\frac{p_{C}(1-q_{C|C})}{1-p_{C}}\right)^{k_{C}}\left(\frac{1-2p_{C}+q_{C|C}p_{C}}{1-p_{C}}\right)^{k-k_{C}}k_{C}(\Pi_{C}^{D}-\Pi_{F}^{D})\nonumber\\
		&-\frac{p_C}{4kN}\sum_{k_C=0}^k\binom k{k_C}q_{C|C}^{k_C}(1-q_{C|C})^{k-k_C}(k-k_C)(\Pi_D^C-\Pi_F^C), \\
		F_{2}(p_{C},q_{C|C})&=\frac{1-q_{C|C}}{4kN}\left[1+(k-1)\frac{p_{C}-q_{C|C}}{1-p_{C}}\right]. 
	\end{align}
\end{subequations}

With $0<s\ll1$, we can reduce the system as ${\dot{p}}_C = sF_1(p_C,\ q_{C|C})$ and $s{\dot{q}}_{C|C} = sF_2(p_C,\ q_{C|C})$. It can be found that $q_{C|C}$ equilibrates much more faster than $p_C$ when $s$ is small and $sF_2(p_C,\ q_{C|C})\neq 0$. Moreover, it may rapidly converge to the root defined by $F_2(p_C,\ q_{C|C})=0$ as time $t \to \infty$. Thus we have
\begin{equation}\label{eq_b12}
	q_{C|C}=\frac{k-2}{k-1}p_C+\frac{1}{k-1}.
\end{equation}

The system can be described only by \( p_C \). Consequently, we have \( q_{D|C}=\frac{k-2}{k-1}(1-p_C) \), \( p_{CD}=p_{DC}=\frac{k-2}{k-1}p_C(1-p_C) \), \( q_{C|D}=\frac{k-2}{k-1}p_C \), \( q_{D|D}=1-\frac{k-2}{k-1}p_C \), and \( p_{DD}=(1-p_C)(1-\frac{k-2}{k-1}p_C) \). At this point, 
\begin{subequations}
	\begin{align}
		\lambda_1&=p(1+\delta)(1+\frac{(k-2)p_C+1}{k-1}\delta)^{k-1},\\
            \lambda_{2}&=p(1+\frac{(k-2)p_{C}}{k-1}\delta)^{k-1},\\
            \lambda_{3}&=(1-p)(1-\delta)(1-\frac{(k-2)p_{C}+1}{k-1}\delta)^{k-1},\\
            \lambda_4&=(1-p)(1-\frac{(k-2)p_C}{k-1}\delta)^{k-1}.
	\end{align}
\end{subequations}

When $N$ is large ($N \to \infty$), the dynamical equations can be further written as
\begin{equation}\label{eq_b13}
\dot{p}_C = s F_1\left(p_C,\ \frac{k-2}{k-1}p_C+\frac{1}{k-1}\right)=\frac{s(k-2)}{4(k-1)}p_C(1-p_C)g(p_C),
\end{equation} 
where
\begin{equation*}
	\begin{aligned}
		g(p_C)&=\phi_1-\phi_2-\phi_3+\phi_4-(k-2)p_C(\phi_2-\phi_4),\\
            \phi_1&=\frac{r}{(k+1)\delta}\{[(k-2)p_C+1][(1+\delta)(\lambda_1-\lambda_2)-(1-\delta)(\lambda_3-\lambda_4)]+(\lambda_1-\lambda_2-\lambda_3+\lambda_4)+\delta(k-1)(\lambda_2+\lambda_4)\}-(k+1),\\
            \phi_2&=\frac{r}{\left(k+1\right)\delta}\left(\lambda_1-\lambda_2-\lambda_3+\lambda_4\right),\\
            \phi_3&=\frac{r}{\left(k+1\right)\delta}(k-2)p_C(\lambda_1-\lambda_2-\lambda_3+\lambda_4)-\delta(k-1)(\lambda_2+\lambda_4)+k+1,\\
            \phi_4&=\frac{r}{\left(k+1\right)\delta}[\left(1+\delta\right)(\lambda_1-\lambda_2)-\left(1-\delta\right)(\lambda_3-\lambda_4)].
	\end{aligned}
\end{equation*}

When \( \delta=0 \), we have \( \lambda_1=\lambda_2=p \) and \( \lambda_3=\lambda_4=1-p \). Further we get \( \phi_1=\frac{r}{(k+1)}[2(k-2)p_C+k+3]-(k+1) \), \( \phi_2=\phi_4=\frac{2r}{\left(k+1\right)} \), and \( \phi_3=\frac{r}{\left(k+1\right)}[2(k-2)p_C-k+1]+k+1 \). Thus, \( g(p_C)=2r-2(k+1) \), and this degeneracy result shows that the success condition for the evolution of cooperation in structured populations is $r > k+1$, which is consistent with $r>G$ for well-mixed populations.

\section{Asymmetric nonlinear coefficients in public goods games.}\label{sup3}

In the analysis of the main text, we use symmetric nonlinear coefficients. In that section, we study the effect of a more general form, i.e., asymmetric nonlinear coefficients, on the evolution of cooperation. In our model, it is assumed that the synergy factor \(\omega_1=1 + \delta_1\) and the discounting factor \(\omega_2=1 - \delta_2\). Let \(\pi_i^j\) represent the payoffs for strategy $i \in \{C,D\}$ participating in a PGG with effect \(j \in \{H,L\}\). Thus, we get
\begin{subequations}
    \begin{align}
        \pi_{C}^{H}(n_{C}) &=\frac{r}{G}\cdot\frac{(1+\delta_1)^{n_{C}}-1}{\delta_1}-1,  \\
        \pi_{C}^{L}(n_{C}) &=\frac{r}{G}\cdot\frac{1-(1-\delta_2)^{n_{C}}}{\delta_2}-1,  \\
        \pi_{D}^{H}(n_{C}) &=\frac{r}{G}\cdot\frac{(1+\delta_1)^{n_{C}}-1}{\delta_1},  \\
        \pi_{D}^{L}(n_{C}) &=\frac{r}{G}\cdot\frac{1-(1-\delta_2)^{n_{C}}}{\delta_2}.
    \end{align} 
\end{subequations}
In the context of a stochastic nonlinear PGG, the payoffs of the cooperators and defectors can be expressed as
\begin{subequations}
    \begin{align}
        \pi_{C}(n_{C})&=p\pi_{C}^{H}(n_{C})+(1-p)\pi_{C}^{L}(n_{C})\nonumber\\&=p\frac{r[(1+\delta_1)^{n_{C}}-1]}{G\delta_1}+(1-p)\frac{r[1-(1-\delta_2)^{n_{C}}]}{G\delta_2}-1,\\
		\pi_{D}(n_{C})&=p\pi_{D}^{H}(n_{C})+(1-p)\pi_{D}^{L}(n_{C})\nonumber\\&=p\frac{r[(1+\delta_1)^{n_{C}}-1]}{G\delta_1}+(1-p)\frac{r[1-(1-\delta_2)^{n_{C}}]}{G\delta_2}.
    \end{align} 
\end{subequations}

\begin{figure*}[htbp]
    \centering
    \includegraphics[width=1\textwidth]{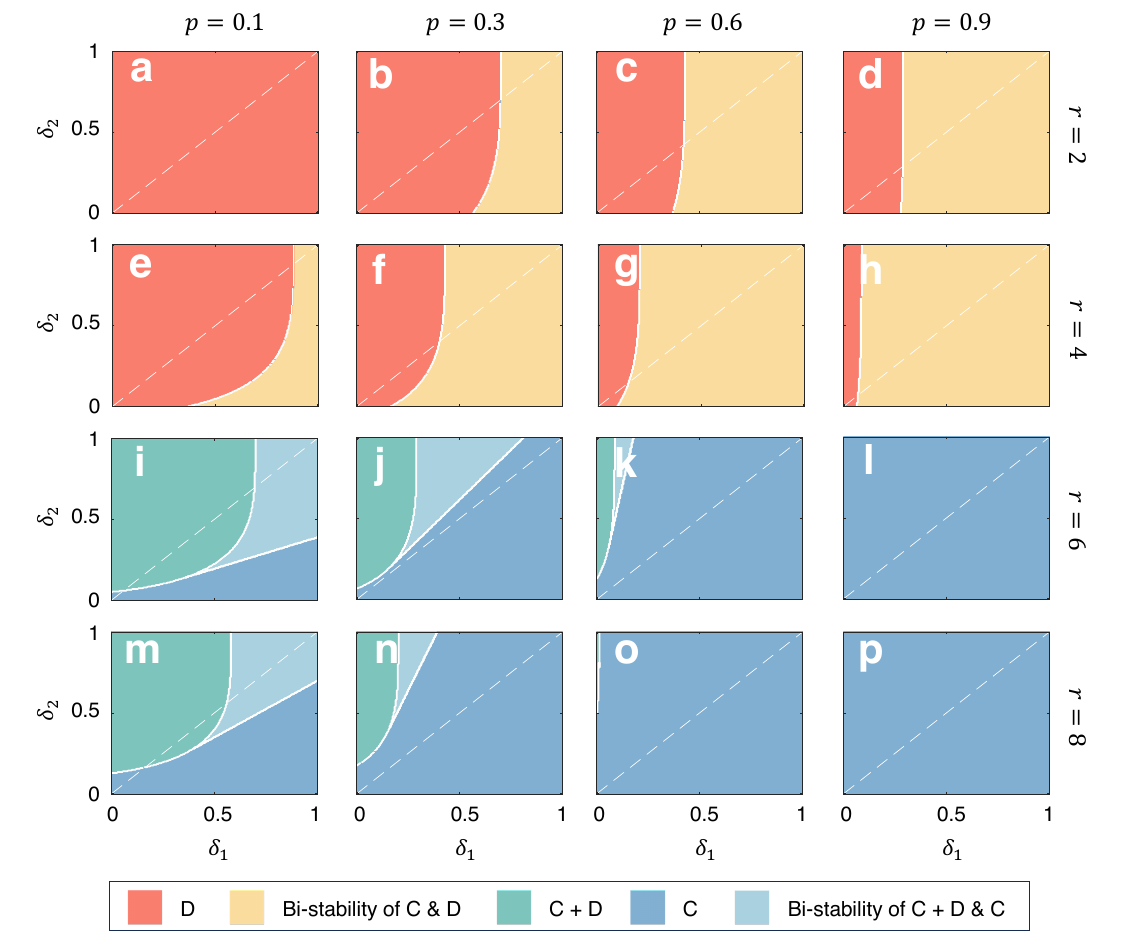}
    \caption{\textbf{Synergy as a major determinant of the evolution of cooperation.} We show the phase diagrams in the $\delta_1$-$\delta_2$ parameter plane when $p\in \{0.1,0.3.0.6,0.9\}$, $r\in \{2,4,6,8\}$ and $G=5$. The phase transition process of the system is mainly controlled by $\delta_1$, especially when the value of $\delta_2$ is large. Whereas, the change of $\delta_2$ causes the system to undergo a phase transition only when the $p$-value is small and within a certain range of $\delta_2$ values. The higher the probability of becoming a synergistic group, the easier it is for synergy factors to lead the evolution of cooperation. Panels \textbf{a}-\textbf{h} denote cases with small values of $r$. From the full parameter plane, an increase in $p$ decreases the parameter region of D and increases the parameter region of Bi-stability of C $\&$ D. Panels \textbf{i-p} denote cases with larger values of $r$. In terms of the full parameter plane, an increase in $p$ decreases the parameter region of C+D and Bi-stability of C+D $\&$ C and increases the parameter region of C. The white dashed line indicates the case where $\delta_1$=$\delta_2$=$\delta$. It can be observed that, except for some extreme cases (i.e., small $p$-values and large $r$-values, corresponding to panels \textbf{i, j, m, n}), the use of a uniform $\delta$ value is able to represent the main evolutionary results of the system.}
	\label{delta1_delta2}
\end{figure*}

In a well-mixed population, the evolutionary dynamics are determined by the average payoff difference between cooperators and defectors. Let $x$ represent the fraction of cooperators and $1-x$ the fraction of defectors. According to Eq.~(\ref{eqs1}), we can get the average payoff for cooperators ${\Pi}_C$ and defectors ${\Pi}_D$ as
\newpage
\begin{subequations}
	\begin{align}
		\Pi_C&=\sum_{i=0}^{G-1}\binom{G-1}{i}x^i(1-x)^{G-1-i}\pi_C(i+1)\nonumber\\
            &=\sum_{i=0}^{G-1}\binom{G-1}ix^i(1-x)^{G-1-i}\left[p\frac{r\left[(1+\delta_1)^{i+1}-1\right]}{G\delta_1}+(1-p)\frac{r\left[1-(1-\delta_2)^{i+1}\right]}{G\delta_2}-1\right]\nonumber\\
            &=\frac rG\cdot\left[\frac p{\delta_1}\left[(1+\delta_1)(1+\delta_1x)^{G-1}-1\right]+\frac{1-p}{\delta_2}\left[1-(1-\delta_2)(1-\delta_2x)^{G-1}\right]\right]-1,\\
		\Pi_D&=\sum_{i=0}^{G-1}\binom{G-1}{i}x^i(1-x)^{G-1-i}\pi_D(i)\nonumber\\
            &=\sum_{i=0}^{G-1}\binom{G-1}ix^i(1-x)^{G-1-i}\left[p\frac{r\left[(1+\delta_1)^{i}-1\right]}{G\delta_1}+(1-p)\frac{r\left[1-(1-\delta_2)^{i}\right]}{G\delta_2}\right]\nonumber\\
            &=\frac rG\cdot\left[\frac p{\delta_1}\left[(1+\delta_1 x)^{G-1}-1\right]+\frac{1-p}{\delta_2}\left[1-(1-\delta_2 x)^{G-1}\right]\right].
		\end{align}
\end{subequations}

Then, the average payoff difference is
\begin{equation}
    \Pi_{C}-\Pi_{D}=\frac{r}{G}[p(1+\delta_1 x)^{G-1}+(1-p)(1-\delta_2 x)^{G-1}]-1.
\end{equation}
The probability that a cooperator transforms into a defector is
\begin{equation}
    P_{C\leftarrow D}=\frac1{1+\exp[s(\Pi_{C}-\Pi_{D})]},
\end{equation}
where $s>0$ denotes the strength of selection.

Eventually, we can get the evolutionary dynamics as
\begin{equation}
	\begin{aligned}
		\dot{x}&=x(1-x)\tanh\left\{\frac{s}{2}\left[ \frac{r}{G}[p(1+\delta_1 x)^{G-1}+(1-p)(1-\delta_2 x)^{G-1}]-1\right]\right\}.
	\end{aligned}
\end{equation}
At this point, the effect of synergy on the evolution of cooperation is reflected in $p(1+\delta_1 x)^{G-1}$, and the effect of discounting on the evolution of cooperation is reflected in $(1-p)(1-\delta_2 x)^{G-1}$. From Fig.~\textbf{S}\ref{delta1_delta2}, it can be found that synergy is able to win the competition with discounting and as a major determinant of the evolution of cooperation.

\section{Synergy and discounting in the aggregation dilemma.}\label{sup4}

In the standard linear PGG, as the size of the group increases, not only will the social dilemma be aggravated (more defectors may free-ride), but there will also be an aggregation dilemma. The emergence of the aggregation dilemma can be understood from two aspects. On the one hand, increasing the size of the group will cause players to participate in more PGGs, thereby increasing potential benefits; on the other hand, it will reduce the benefits generated by the unit contribution of the cooperators, making them more dependent on the strategies of others and increasing risks~\cite{SM_aggregation}.

\begin{figure*}[!ht]
    \centering
    \includegraphics[width=1\textwidth]{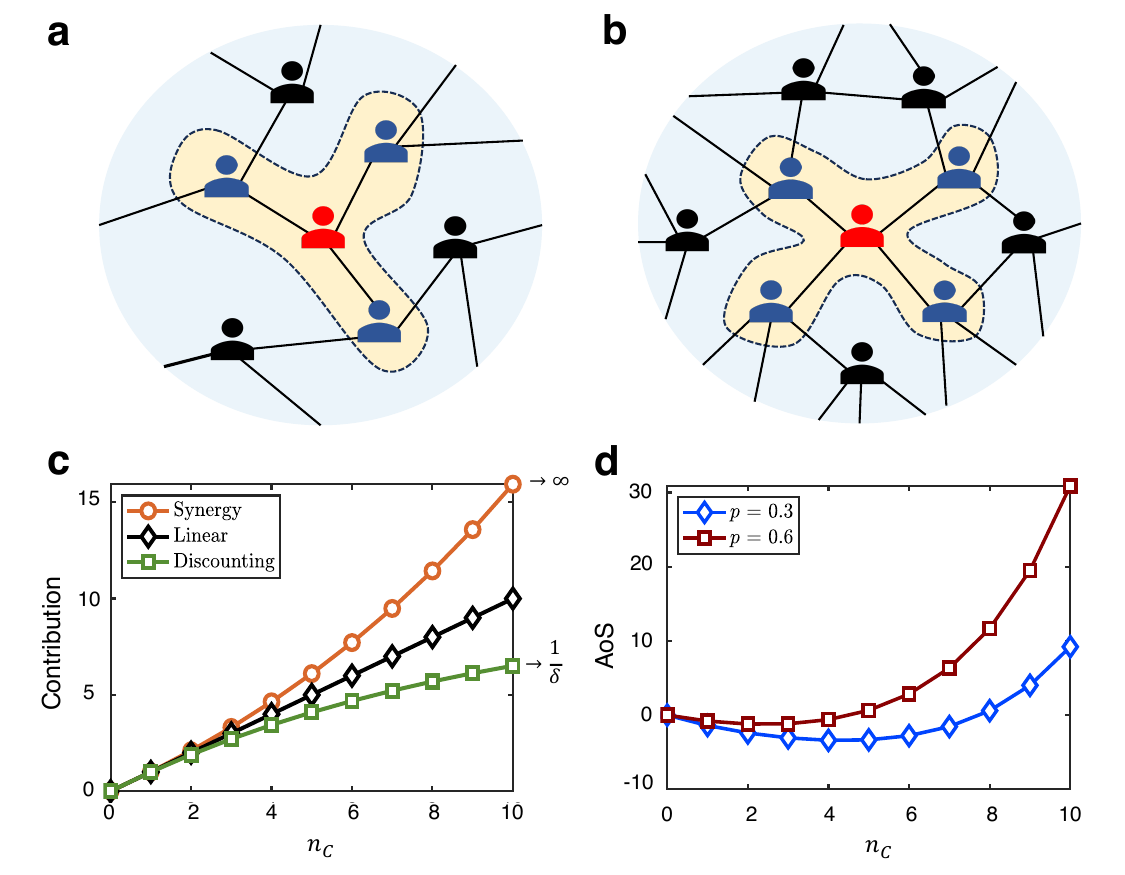}
    \caption{\textbf{Synergy can solve the aggregation dilemma.} Panels \textbf{a} and \textbf{b} show player configurations for different group sizes. An increase in group size leads to an aggregation of players, which on the one hand increases the potential payoffs due to the possibility of participating in more PGGs, and on the other hand increases the risk due to a decrease in the payoffs generated by unit contributions. Panel \textbf{c} reveals the relationship between the number of cooperators $n_C$ and contributions under three different interactions (synergy, linear and discounting). As $n_C$ increases, the difference between the contributions generated by the two nonlinear interactions in the same $n_C$ condition gradually increases compared to the linear case. When $n_C$ is infinite, the contribution produced by the synergy tends to $\infty$, while the contribution produced by the discounting tends to $1/ \delta$. Panel \textbf{d} shows the advantage of synergy (AoS) in stochastic nonlinear PGG. Moderate $n_C$ is most unfavorable for synergy and the intensity of the aggregation dilemma is strongest at this point, thus potentially hindering the evolution of cooperation.}
	\label{competition_nc}
\end{figure*}

As shown in Fig.~\textbf{S}\ref{competition_nc}\textbf{a} and \textbf{b}, an increase in group size by 1 will cause the player to participate in PGG one more time in the same round (4 PGGs in panel \textbf{a}, 5 PGGs in panel \textbf{b}), and also reduce the benefits generated by the contributions of the cooperators (from $r/4$ in panel \textbf{a} to $r/5$ in panel \textbf{b}). With a larger group size, the risk of achieving high benefits will be greater, especially at this time, the increase in social dilemmas will reduce the willingness of players to cooperate.

In our model, synergistic interactions can overcome the aggregation dilemma, while discounting interactions can exacerbate the aggregation dilemma. According to Eq.~(\ref{eq_1}) in the main text, we can define the contributions of participants in the public pool under linear as $n_C$, and the contribution of synergy and discounting as respectively
\begin{subequations}
    \begin{align}
        &\mathrm{Synergy}=\frac{(1+\delta)^{n_{C}}-1}{\delta},  \\
        &\mathrm{Discounting}=\frac{1-(1-\delta)^{n_{C}}}{\delta}.
    \end{align}
\end{subequations}
An increase in group size implies an increase in the number of cooperators that can be accommodated. For deterministic nonlinear PGG, it can be found from Fig.~\textbf{S}\ref{competition_nc}\textbf{c} that as the number of cooperators in the group increases, the increase in the contribution of synergy relative to linear versus the decrease in the contribution of discounting relative to linear shows an asymmetric relationship. When $n_C$ is small, there is little difference in the overall contributions generated by the three different interaction modes. And as $n_C$ increases, the increase in contribution from synergy will be much greater than the decrease in contribution from discounting. Thus, for deterministic nonlinear PGG, as the group size increases, synergistic interactions help overcome the aggregation dilemma and discounted interactions exacerbate the aggregation dilemma. 

When considering stochastic nonlinear PGG, both synergy and discounting will exist in the population. In this case, define a metric to measure the advantage of synergy (AoS), namely
\begin{equation}
    \mathrm{AoS}=p\vert \frac{(1+\delta)^{n_{C}}-1}{\delta} - n_C \vert - (1-p)\vert \frac{1-(1-\delta)^{n_{C}}}{\delta}-n_C \vert.
\end{equation}

As can be seen from Fig.~\textbf{S}\ref{competition_nc}\textbf{d}, as the number of cooperators in the group increases, the advantage of synergy undergoes a process of first decreasing and then increasing. In other words, there is a moderate $n_C$ that makes the advantage of synergy reach the minimum (at this time, the value is less than 0, indicating that the discounting is dominant). Therefore, for stochastic nonlinear PGG, moderate group size will aggravate the aggregation dilemma, while synergistic interaction at a large scale helps to overcome the aggregation dilemma. When we also take the social dilemma into account, moderate group size will show poor cooperative evolution results, and increasing the group size can promote the evolution of cooperation (see Fig.~\ref{hz_Gk} in the main text).

\section{Monte Carlo simulation to validate evolutionary dynamics.}\label{sup5}
In this section we perform simulations in well-mixed populations to validate our evolutionary results in Fig.~\ref{dynamic} in the main text.

In a well-mixed population with $N$ players, each player can interact with all other players. In each public goods game consisting of $G$ players, the other $G-1$ co-players are randomly selected in the population (each player has the same probability of being selected). If the current group is a synergy group, $w = 1 + \delta$, and if it is a discounting group, $w = 1 - \delta$. Then the payoff can be computed by Eq.~(\ref{eq_1}) in the main text. The strategy updating process is simulated using asynchronous Monte Carlo simulation, where at each standard Monte Carlo step (MCS), the replacement event occurs only once, a randomly selected player $i$ chooses a neighbor $j$ at random and imitates $j$'s strategy according to 
\begin{equation}
	P_{i\leftarrow j}=\frac1{1+\exp[s(\Pi_i-\Pi_j)]},
\end{equation}
where $s$ denotes the strength of selection~\cite{SM_fermi}, $\Pi_i$ ($\Pi_j$) denotes the total payoffs of player $i$ ($j$).

\begin{figure*}[ht]
    \centering
    \includegraphics[width=1\textwidth]{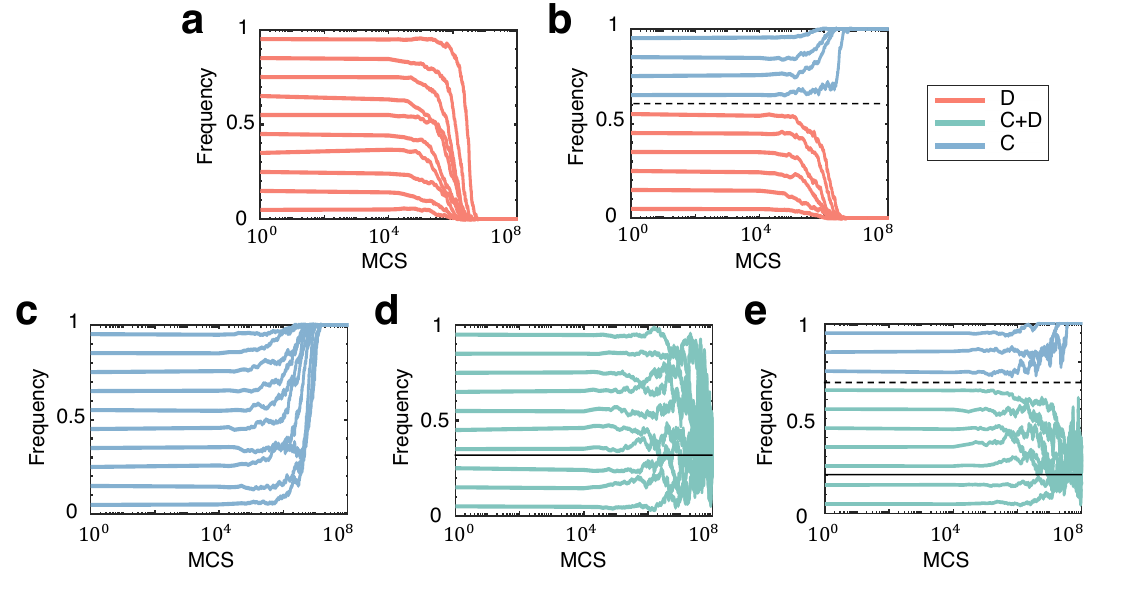}
    \caption{\textbf{Simulation results in well-mixed populations.} In each panel, we perform simulations for different initial states with the same parameters to determine all possible evolutionary outcomes for the population. At each run, the population will go through $1 \times 10^8$ MCSs to ensure that the system reaches stability. For a given initial proportion of cooperators, the population evolves to a total of three stable states, i.e., full defection (D), coexistence of defection and cooperation (D+C) and full cooperation (C). We verify this deterministic state with 20 independent runs for each particular initial setup. Panel \textbf{a} shows that only one state of full defection exists for the population. Panel \textbf{b} emerges in two states, full cooperation and full defection. Panel \textbf{c} shows that only one state of full cooperation exists for the population. Panel \textbf{d} shows that cooperation and defection coexist in the population. Panel \textbf{e} shows the two states of cooperation and defection coexisting as well as full cooperation, The solid and dashed lines correspond to stable and unstable internal equilibrium points in the evolutionary dynamics, respectively. The parameters are set as $N=1 \times 10^4$, $s=0.1$, $G=5$, \textbf{a} $r=3$, $p=0.6$, $\delta=0.2$, \textbf{b} $r=3$, $p=0.6$, $\delta=0.4$, \textbf{c} $r=5.1$, $p=0.6$, $\delta=0.1$, \textbf{d} $r=5.1$, $p=0.4$, $\delta=0.1$, \textbf{e} $r=5.1$, $p=0.4$, $\delta=0.15$.}\label{simu_time}
\end{figure*}

We validate five evolutionary states in which populations can emerge. Fig.~\textbf{S}\ref{simu_time} uses the same parameters as in Fig.~\ref{dynamic} in the main text. In the simulation results, five states of the population are consistent with the theoretical analysis. In addition, the boundary point of the initial frequency of cooperators corresponding to the bi-stable state is consistent with the unstable internal equilibrium point in the theoretical analysis (as shown by the dotted line). The frequency of cooperators when cooperators and defectors coexist is consistent with the stable internal equilibrium in the theoretical analysis (as shown in the straight line).

\end{document}